\documentclass[runningheads]{llncs}
\usepackage{graphicx}

\usepackage{amsmath,amssymb,amsfonts}
\usepackage{pifont}
\usepackage{listings}
\usepackage{array}
\usepackage{multirow}
\usepackage{tabularx}
\usepackage{hyperref}

\usepackage{wrapfig}

\newcommand{\Analyser}{Guardian}

\let\oldding\ding% Store old \ding in \oldding
\renewcommand{\ding}[2][1]{\scalebox{#1}{\oldding{#2}}}%

\newcounter{lstannotation}
\renewcommand{\thelstannotation}{\ding[1.1]{\number\numexpr171+\arabic{lstannotation}}}
\newcommand{\lannotation}[1]{\refstepcounter{lstannotation}\label{#1}\thelstannotation}

\newcommand{\entry}{\textit{entry}}
\newcommand{\secure}{\textit{secure}}
\newcommand{\ocall}{\textit{ocall}}
\newcommand{\exit}{\textit{exit}}

\begin{document}
\title{Guardian: symbolic validation of orderliness in {SGX} enclaves}

\author{Pedro Antonino\inst{1} \and
	Wojciech Aleksander Wo\l{}oszyn\inst{1,2} \and
	A. W. Roscoe \inst{1,3,4}}
	
\authorrunning{P. Antonino et al.}
\institute{The Blockhouse Technology Limited, Oxford, UK \and 
	Mathematical Institute, University of Oxford, Oxford, UK \and
	University College Oxford Blockchain Research Centre, Oxford, UK \and
	Department of Computer Science, University of Oxford, Oxford, UK\\
	\email{\{pedro, wojciech\}@tbtl.com, awroscoe@gmail.com}}

\maketitle

\begin{abstract}
% TEEs and Intel SGX.
Modern processors can offer hardware primitives that allow a process to run in isolation. These primitives implement a trusted execution environment (TEE) in which a program can run such that the integrity and confidentiality of its execution are guaranteed. Intel's Software Guard eXtensions (SGX) is an example of such primitives and its isolated processes are called \emph{enclaves}. These guarantees, however, can be easily thwarted if the enclave has not been properly designed. Its interface with the untrusted software stack is arguably the largest attack surface that adversaries can exploit; unintended interactions with untrusted code can expose the enclave to memory corruption attacks, for instance. In this paper, we propose a notion of an \emph{orderly} enclave which splits its behaviour into several execution phases each of which imposes a set of restrictions on accesses to untrusted memory, phase transitions and registers sanitisation. A violation to these restrictions indicates an undesired behaviour which could be harnessed to perpetrate attacks against the enclave. We also introduce \Analyser{}: a tool that uses symbolic execution to carry out the validation of an enclave against our notion of an orderly enclave; in this process, it also looks for some typical memory-corruption vulnerabilities. We discuss how our approach can prevent and flag enclave vulnerabilities that have been identified in the literature. Moreover, we have evaluated how our approach fares in the analysis of some practical enclaves. \Analyser{} was able to identify real vulnerabilities on these enclaves which have been acknowledged and fixed by their maintainers.

\end{abstract}

\section{Introduction}
\label{sec:intro}

% TEE and SGX
Some modern processors offer hardware primitives that implement trusted execution environments (TEE)~\cite{Maene18}. When (part of) a process is run inside a TEE, its execution's integrity and confidentiality are preserved. The use of a TEE essentially splits computing resources into two categories: trusted and untrusted. While trusted resources are managed by the TEE and, therefore, enjoy its protection and guarantees, untrusted resources lie outside its trust boundary. Intel's TEE implementation, called Software Guard eXtensions (SGX)~\cite{SGXSDK,Costan16}, allows a process to have a protected memory range that is called an \emph{enclave}. 
Enclaves are isolated from other applications, the operating system, hypervisors, and even other enclaves. 
% Problems with TEEs and SGX
The protection offered by TEEs, however, can be thwarted by a careless enclave design and implementation. The interface between enclave and the untrusted software stack, especially, has to be very carefully designed to prevent attacks. There is little point in running a program inside a TEE if the untrusted software stack can manipulate its execution and breach its guarantees. As enclaves are being rapidly adopted~\cite{Kust20,Cheng19,Arnautov16,Schuster15,Priebe18},  frameworks to analyse and improve on their security are increasingly needed.

% Our solution challenges
We propose the notion of an \emph{orderly enclave} which splits the execution of enclaves into phases: \entry{}, \secure{}, \ocall{}, and \exit{}, each of which imposes a set of restrictions on the behaviour of the enclave in the form of untrusted-memory-access policies and sanitisation conditions. For instance, in the \entry{} phase---which corresponds to the point when the untrusted host application has started the enclave and handed control over to it---we require CPU registers, which are shared between the untrusted hosted application and enclave, to be sanitised before the \secure{} phase starts to prevent untrusted register values from being used in \emph{secure} computations. Violations to these policies and conditions represent flaws in the design or implementation of the enclave, or even underlying \emph{vulnerabilities} which might be used to breach enclave's guarantees. 

We also propose \emph{\Analyser{}}: a tool that uses symbolic execution to validate whether an enclave binary is orderly---it also tries to detect typical memory-corruption vulnerabilities in the process. Symbolic execution approaches are very useful for the analysis of programs, especially when targeting intermediate languages or bytecode \cite{Cadar08,Shoshitaishvili16,Lindner18,Mossberg19,Saudel15,Puasuareanu13}. They analyse the execution of a program based on symbolic representatives of its variables. This symbolic exploration tends to cover a large number of code paths, making it an effective strategy for finding bugs and vulnerabilities in programs. We use \emph{angr}~\cite{Shoshitaishvili16}---a symbolic execution engine that focuses on usability---as a backend for our tool. The developer of the enclave can use our tool in an iterative process where it validates design and implementation, fixing violations until no more of them are found.

% Fine described contributions as solutions to these challenges:
Our contributions are as follows. The definition of an orderly enclave which captures a helpful discipline for the developer with respect to interactions between enclave and the untrusted software stack. We also give a practical overview of how our framework should be applied to typical Intel SGX SDK enclaves, detailing how the SDK meets the restrictions imposed by each execution phase of an orderly enclave. Furthermore, we propose a tool that uses symbolic execution to validate an enclave binary against our notion of an orderly enclave, while also identifying common memory-corruption vulnerabilities in the process. Our definition and tool was inspired by~\cite{Bulck19}. While that paper manually identifies a number of enclave vulnerabilities, our notion of an orderly enclave and \Analyser{} \emph{automatically} identifies such vulnerabilities or their effects. We present a practical evaluation of our framework where we show how it fares in analysing some real enclaves. We identified issues in enclaves which went undetected when analysed by other tools. These results demonstrate that our framework and tool complement similar state-of-the-art approaches. These issues were acknowledged and fixed by the enclave maintainers.

\noindent
\textbf{Outline.} Section~\ref{sec:related} examines related work, and Section~\ref{sec:background} introduces the concepts related to Intel SGX, symbolic execution, and \emph{angr}. Section~\ref{sec:analysis} presents our framework, discusses how it identifies some typical enclave vulnerabilities, and presents an evaluation on how it fares when applied to some real enclaves. We present our concluding remarks in Section~\ref{sec:conclusion}.

\section{Related Work}
\label{sec:related}

%Biondo and Lee: Memory corruption attacks: 
The identification of enclave vulnerabilities has been the focus of many recent papers. For brevity, in this section, we focus on the work that we believe to be most closely related to ours.

Lee at al. present the first memory-corruption attack against Intel SGX enclaves in~\cite{Lee17}. The \emph{Dark-ROP} method finds attack gadgets in the enclave's code, which are later explored to execute malicious code that can both exfiltrate enclave's derived keys and generate attacker-controlled reports.  Biodo et al.~\cite{Biondo18} improve on this initial attack technique by proposing two new attacks that can be carried out under weaker adversarial assumptions; it works for enclaves with a randomised memory layout (using, for instance, SGX-Shield~\cite{Seo17}) and do not require kernel privileges. These papers highlight the need for tools such as \Analyser{} to detect vulnerable memory manipulations.

% Tale of two worlds: focus on runtime libraries + no automated analysis + no definition of well-defined enclave
Bulck at al.~\cite{Bulck19} manually analyse runtime libraries for several TEE implementations and how poorly designed interfaces between trusted and untrusted code can lead to vulnerabilities. It considers Intel SGX, Keystone~\cite{Lee20}, and Sanctus~\cite{Noorman17}, as far as TEE implementations, and runtime libraries including Intel SGX SDK and Open Enclave SDK~\cite{OpenEnclave}.  It identifies a number of exploitable vulnerabilities that can be used to perpetrate powerful attacks against enclaves. That work~\cite{Bulck19} inspired our paper. While that paper identifies a number of vulnerabilities manually, \Analyser{} implements machine-testable conditions described in our notion of an \emph{orderly enclave}, giving rise to a framework that automatically looks for such vulnerabilities.

% TeeRex: no notion of well-design enclave + no runtime analysis + no ocalls analysis + debug symbols
\emph{TeeRex}~\cite{Cloosters20} is a tool that relies on symbolic execution, using \emph{angr}, to find memory corruption vulnerabilities in Intel SGX enclaves. TeeRex is arguably the work that is closest to ours. Both approaches use symbolic execution to look for vulnerabilities, but they are based on fundamentally different ideas. While TeeRex focuses on finding typical exploit primitives within enclaves, we look for violations of policies to access untrusted memory and sanitisation conditions, both of which are part of our \emph{orderly enclave} definition. Moreover, we analyse the trusted execution spanning from runtime library to user code, whereas TeeRex focuses on user-defined code. It has been shown that runtime-library code can also be flawed~\cite{Bulck19}. Anecdotally, the binaries analysed by TeeRex in~\cite{Cloosters20} are built with an Intel SDK version that exhibits register sanitisation violations, typically afflicting runtime-library code. Our framework is able to identify such vulnerabilities whereas TeeRex is not. Our approach requires a more detailed input from the user in the form of annotations of specific program points, whereas TeeRex only needs addresses of some specific enclave functions. Both \Analyser{} and TeeRex have heuristics to automatically identify these addresses but TeeRex does not use debug symbols and so it can be applied to closed source binaries, unlike \Analyser{}. In Section~\ref{sec:evaluation}, we present violations to our orderly-enclave definitions that were detected by our framework in some practical enclaves but which went undetected when analysed by TeeRex. They have been acknowledged and fixed by the corresponding enclave maintainers.

% Bounds checking
\emph{SGXBounds}~\cite{Kuvaiskii17} is a framework that instruments enclaves with extra boundary metadata about different memory segments (heap, stack, global objects) so that out-of-boundary violations can be identified and handled at runtime. It adapts ideas of memory-safety-hardening paradigms for traditional programs to SGX enclaves~\cite{Oleksenko18,Serebryany12}. Even though both this work and our own are concerned with some form of boundary checking, they differ radically in intent and approach. While SGXBounds detects typical memory-corruption attacks that take place within the enclave, such as buffer-overflows involving the trusted stack and heap, our framework tries to identify vulnerabilities in the new attack surface arising from the interactions between trusted and untrusted world. 

% Side channel attacks
Many side-channel attacks against enclaves have been identified in recent papers~\cite{Gotzfried17,Bulck18,Bulck18b,Bulck18c,Chen19,Schaik19,Murdock20,Schaik20,Alder20,Ragab21,Chen21}; SGX has not been designed to prevent them~\cite{SGXSDK}. There are attacks using various kinds of side-channel oracles---power-based, timing-based, cache-based, FPU-based, etc.---to exfiltrate secrets and influence the behaviour of enclaves. Symbolic execution can be harnessed to identify also side-channel oracle primitives. For instance, alignment faults can be harnessed to create a side-chain oracle~\cite{Bulck19}, and by enforcing that the AC flag is cleared upon enclave entry, \Analyser{} can identify whether the enclave's code exhibits the behaviour necessary to create such an oracle.

\section{Background}
\label{sec:background}

This section introduces the concepts about TEEs and symbolic execution necessary to make our paper self-contained.

\subsection{Intel SGX}
\label{sec:tee}

%General TEEs
Intel's TEE implementation, called Software Guard eXtensions (SGX)~\cite{SGXSDM,Costan16}, allows an untrusted host process to define a range in its virtual memory where integrity-protected and confidential code and data are hosted; this isolated memory is called an \emph{enclave}. SGX extends Intel's traditional instruction set with privileged instructions\footnote{In fact, leaf functions to instruction \lstinline|encls|.}---such as \lstinline|ecreate|, \lstinline|eadd|, \lstinline|eextent|, \lstinline|einit| and \lstinline|eremove|---to create, initialise, and dispose of this protected memory range. The confidentiality, authenticity, and freshness of the memory in this range is guaranteed by a combination of SGX privileged instructions, which protect swapped pages, and Intel SGX's Memory Encryption Engine~\cite{Gueron16}, which protect pages in physical memory. SGX also introduces unprivileged/user instructions\footnote{Leaf functions to instruction \lstinline|enclu|.}  that can be used to execute enclave code. User code can enter and leave the enclave's code using instructions \lstinline|eenter| and \lstinline|eexit|, respectively. Aside from these synchronous enclave transitions, the enclave can also be interrupted and resumed. 

% Intel SDK and edger8r
Built upon these hardware primitives, Intel offers an SDK to help developers create enclaves using C/C++~\cite{SGXSDK,SGXSDKRef,SGXSDKRep}. The enclave programming model relies on two key concepts: an \emph{ecall} is a mechanism by which untrusted code can invoke enclave code, whereas an \emph{ocall} is used to call untrusted code from the enclave. The ecall abstraction is implemented by three functions: an \emph{untrusted proxy function}, a \emph{trusted bridge function}, and a \emph{trusted user-defined ecall function}. The untrusted proxy function transparently abstracts the ecall mechanism for the untrusted application. It switches into enclave mode and effectively calls the corresponding bridge function. The bridge function sanitises and manipulates inputs and, in turn, calls the user-defined ecall function, which carries out some trusted computation. The ocall mechanism is implemented similarly but the direction is reversed: proxy function is part of the trusted code whereas bridge and user-defined functions are untrusted. As we detail in Section~\ref{sec:design}, the SDK introduces an approach to automatically generate proxies and bridges from annotated function signatures~\cite{SGXSDKRef} as an attempt to mitigate attacks harnessing improper interactions between trusted and untrusted worlds. Note that while the enclave can read and write to the memory of the host application, the application is unable to directly access the enclave's memory. Alternative TEE implementations include TrustZone~\cite{Pinto19}, SEV~\cite{sev20}, Keystone~\cite{Lee20}, and Sanctus~\cite{Noorman17}.

\subsection{Symbolic execution and \emph{angr}}
\label{sec:sex}

% Symbolic execution
Symbolic execution is a technique that explores a program's state space using symbolic values instead of concrete ones~\cite{King76,Cadar08,Cadar08b,Schwartz10,Cadar11,Puasuareanu13,Saudel15,Avgerinos16,Shoshitaishvili16,Machiry17,Lindner18,Mossberg19}---a symbolic value of a given type denotes \emph{any} possible value of that type. For each command executed, this technique creates a corresponding constraint over these values to capture the command's effects. The constraints accumulated at different execution points represent \emph{symbolic states} that compactly capture a (possibly very large) set of concrete states---in fact, they normally capture all concrete states reaching that point. Branching on a symbolic value leads the execution to be split into two states, each of which capturing one of the branches. The branching conditions are tracked into what is called a \emph{path condition}---a conjunction of all conditions for the branches taken up to that point; this constraint is also part of the symbolic state. A constraint solver is used to check symbolic states for satisfiability---unsatisfiable states are unreachable---and to generate concrete values for the program's variables at the corresponding execution point; these concrete values can be very useful in generating concrete inputs to exercise that execution path.
% angr
\emph{angr}~\cite{Shoshitaishvili16} is a symbolic execution engine that targets binaries. It is a powerful tool that offers a number of built-in analyses and instrumentation constructs while being user-friendly. It has been successfully used to analyse real-world binary code~\cite{Duan19,Redini17} and even enclaves~\cite{Chen19,Cloosters20,Machiry17}. \emph{angr} provides the ability to instrument binaries with breakpoints, which can intercept the execution of specific instructions, and high-level function summaries, which can simulate and add extra behaviour to the binary. These two constructs are fundamental in designing our tool as we detail in Section~\ref{sec:analyser}.

Symbolic execution is an approach that tends to offer a good compromise between coverage and efficiency. Typically, it is slower but offers better coverage than testing or fuzzing, and it offers worse coverage but is quicker than formal verification approaches. We believe that this balance favours the use of symbolic execution. Our tool is intended to be used before enclaves are distributed; it symbolically executes enclaves to report vulnerabilities (i.e. violations to our notion of an orderly enclave) which need to be addressed by the enclave developer.

\section{Analysing orderly Intel SGX enclaves}
\label{sec:analysis}

%General introduction about our approach.
The guarantees offered by TEEs can be easily thwarted if the design and implementation of an enclave is flawed. Confidentiality is pointless if the untrusted host application can just exfiltrate secrets, and isolated execution useless if the adversary can overly influence it. In this section, we propose a design for \emph{orderly enclaves} that splits enclave execution into several phases, each of which has a specific purpose and expectations for the code being executed there. Building upon this design definition, we also discuss how \Analyser{} analyses binaries, looking for violations of our design and potentially uncovering vulnerabilities.

SGX was designed with a very powerful adversary in mind. The trusted computing base of an SGX enclave consists of its code, the CPU package, and a few privileged containers~\cite{Costan16} such as architectural enclaves~\cite{SGXSDKRef}. All other elements are considered to be under the control of the attacker. So, privileged system software (hypervisor, OS, firmware, etc), and applications---including the enclave's host application---both code and data, are untrusted and assumed to be under the control of an attacker. Even enclaves are isolated from one another. Intel SGX's threat model does not encompass side-channel attacks; the enclave's developer is in charge of protecting its execution against them~\cite{SGXSDK}. Of course, since untrusted code is in charge of loading and executing the enclave code, there is no guarantee that the enclave's code will be executed but only that its execution cannot be tampered with, without detection, or inspected in the clear. For our purposes, we use the following adversary model.

\begin{definition} The enclave's execution, code and data, is integrity protected and confidential---hence, assumed to be executed as prescribed---but any non-enclave code or data is under the attacker's control and can be freely manipulated. Therefore, values read from or written to non-enclave memory, or executing non-enclave code should account for arbitrary adversarial influence.
\end{definition}

\subsection{Orderly enclaves}
\label{sec:design}

The execution of an \emph{orderly enclave} spans across the following phases: \entry{}, \secure{}, \ocall{}, and \exit{}. Roughly speaking, the \entry{} sanitises the inputs of some secure function, \secure{} carries out the computation of this secure function, \ocall{} is a phase by which the secure function might call into untrusted code---typically to realise some OS-assisted functionality---and \exit{} sanitises the outputs of the secure function. We capture these different phases via annotations $\textit{TransitionAnnotations} = (\textit{Entry}, \allowbreak \textit{Secure}, \textit{OCall}, \textit{Exit})$. The elements in this 4-tuple identify different addresses in the enclave binary denoting transitions between these phases. We discuss these annotations, what each of these phases entails in terms of the enclave's expected behaviour, and how this is achieved by enclaves built with the Intel SGX SDK framework~\cite{SGXSDK,SGXSDKRef} in the following. The SDK code is available at~\cite{SGXSDKRep}; we discuss and target version 2.12 for Linux, unless we state otherwise.

% Entering: clearing of some general purpose and special purpose flags, set some flags, switch to trusted stack.
The pair $\textit{Entry} =  \allowbreak (\textit{EntryAddress}, \allowbreak EntrySanitisationDone)$ annotates the enclave's entry point and the address at which the CPU registers sanitisation must have been completed, respectively. An orderly enclave starts its execution at \textit{EntryAddress} in phase \entry{}, which ends as the \secure{} or \exit{} phase begins. While the transition to \secure{} represents the standard enclave behaviour, transitioning into \exit{} can be caused by some failed validation at \entry{}, for instance. During \entry{}, the enclave code has to set up its state appropriately so that it can securely execute without allowing untrusted resources to overly influence its behaviour. This phase is responsible for: setting up the low-level machine state and securing input arguments over which the enclave will compute. 

The enclave has to properly sanitise CPU registers upon entry as they are shared between trusted and untrusted executions. In the Intel SGX SDK, the \emph{trusted runtime system} (\emph{tRTS}) is the SDK runtime library in charge of such sanitisation---the assembly routine \lstinline{enclave_entry} performs this task. While registers used to pass arguments to the enclave execution must have their values preserved, the other registers which might impact the execution of the enclave have to be treated carefully. For x86-64 architecture, this routine expects registers \emph{rax}, \emph{rbx}, \emph{rcx}, \emph{rdi}, and \emph{rsi} to store input arguments for the enclave's execution. For instance, \emph{rbx} identifies the enclave thread to be executed, \emph{rdi} the index of the ecall to be executed, and \emph{rsi} a pointer to the marshalling structure containing the parameters of the ecall. Other general purpose registers, such as \emph{rdx} and \emph{r8}-\emph{r15}, are cleared by tRTS so that they do not adversely impact the execution of the enclave. The CPU flags also need to be properly handled. For instance, the work in~\cite{Bulck19} shows that the Alignment Check (AC) flag and the Direction Flag (DF)---both part of the \emph{rflags} register---can be harnessed to carry out powerful attacks against enclaves. The SDK clears both flags upon entry. Finally, the enclave should use its own private and protected stack for trusted execution. This phase should make sure that the low-level machine state switches from the untrusted stack to the trusted enclave stack. This routine sets up registers \emph{rbp} and \emph{rsp} so that they point to the trusted stack. We capture this low-level sanitisation with predicate \textit{EntrySanitisationValidation} which holds if and only if all the following conditions are true: registers \emph{rdx}, and \emph{r8}-\emph{r15} are zeroed; \emph{rbp} and \emph{rsp} point to the trusted stack; and flags AC and DF are cleared. We do not constrain the values of \emph{rax}, \emph{rbx}, \emph{rcx}, \emph{rdi}, \emph{rsi} as they store arguments for the enclave execution. So, for an orderly enclave, when address \textit{EntrySanitisationDone} is reached, we specify that this predicate must hold.

During \entry{}, the enclave should also copy input arguments from untrusted into trusted memory. This copy should prevent, for instance, time-of-check time-of-use attacks whereby untrusted memory can be manipulated to bypass checks executed by the enclave. Note that in the adversary model we propose, the attacker is fully in control of the untrusted memory and could alter some values in memory between the enclave checking and using them. The SDK advocates the use of the \emph{Enclave Definition Language} (EDL) and the \emph{edger8r tool}~\cite{SGXSDK,SGXSDKRef} to carry out this copy.
EDL allows developers to annotate pointer parameters with specific copy and validation policies, whereas the edger8r tool generates the code to carry out the copying and validation based on these policies. The \lstinline|in| and \lstinline|out| EDL annotations are called \emph{pointer direction attributes}: they represent whether the parameter is passed from the calling to the called procedure or the other-way around, respectively. Listing~\ref{lst:edl} presents a very basic EDL definition with an ecall and an ocall function. In Listing~\ref{lst:sdk-enclave}, we present abstract templates for prominent functions in the Intel SGX SDK's programming model. While \lstinline|sgx_ecall_function| and \lstinline|ocall_function| represent a bridge ecall function and a proxy ocall function, respectively, which are generated by edger8r based on our EDL definition, \lstinline|ecall_function| is user-defined. Each number in these templates denotes a block of code; they are accompanied by a brief description of their purpose. Function \lstinline|sgx_ecall_function| manipulates the parameters and calls the secure computation denoted by \lstinline|ecall_function|.\footnote{In fact, \lstinline|sgx_ecall_function| receives a \emph{marshalling structure} which contains as elements the values of \lstinline|i| and \lstinline|retv|. For the sake of presentation, we simplify our function definition to use these elements directly. We make a similar simplification for \lstinline|ocall_function|.} The \lstinline|[in]| annotation for \lstinline|i| generates both a pointer validation in \ref{lst:sgx-ecall-check} that checks whether \lstinline|i| points to an \lstinline|int| that lies in the memory outside of the enclave, and the code to copy the \lstinline|int| pointed by \lstinline|i| into the trusted heap in \ref{lst:sgx-ecall-copy-in}; local variable \lstinline|t_i| points to this new trusted \lstinline|int|. During the \entry{} phase, we require an orderly enclave to respect the \textit{EntryPolicy}: the enclave can read from untrusted memory (to carry out argument copies, for instance) but it is not allowed to write to untrusted memory.

\begin{figure}[t]
\begin{minipage}[!t]{0.48\textwidth}
\begin{minipage}{\textwidth}
\begin{lstlisting}[basicstyle=\scriptsize\sffamily,linewidth=\columnwidth,breaklines=true, caption=EDL example., label=lst:edl]
enclave {
trusted { \\ ecalls definintion block
	public void ecall_function([in] int *i, [out] int* retv);
};
untrusted { \\ ocalls definition block
	void ocall_function([in, out] int *j);
};
};
\end{lstlisting}
\end{minipage}
\begin{minipage}{\textwidth}
		\centering
		\caption{Typical execution with our framework's relevant address annotations and their corresponding Intel SGX SDK enclave commands.}
		\label{fig:execution}
		\includegraphics[width=\textwidth]{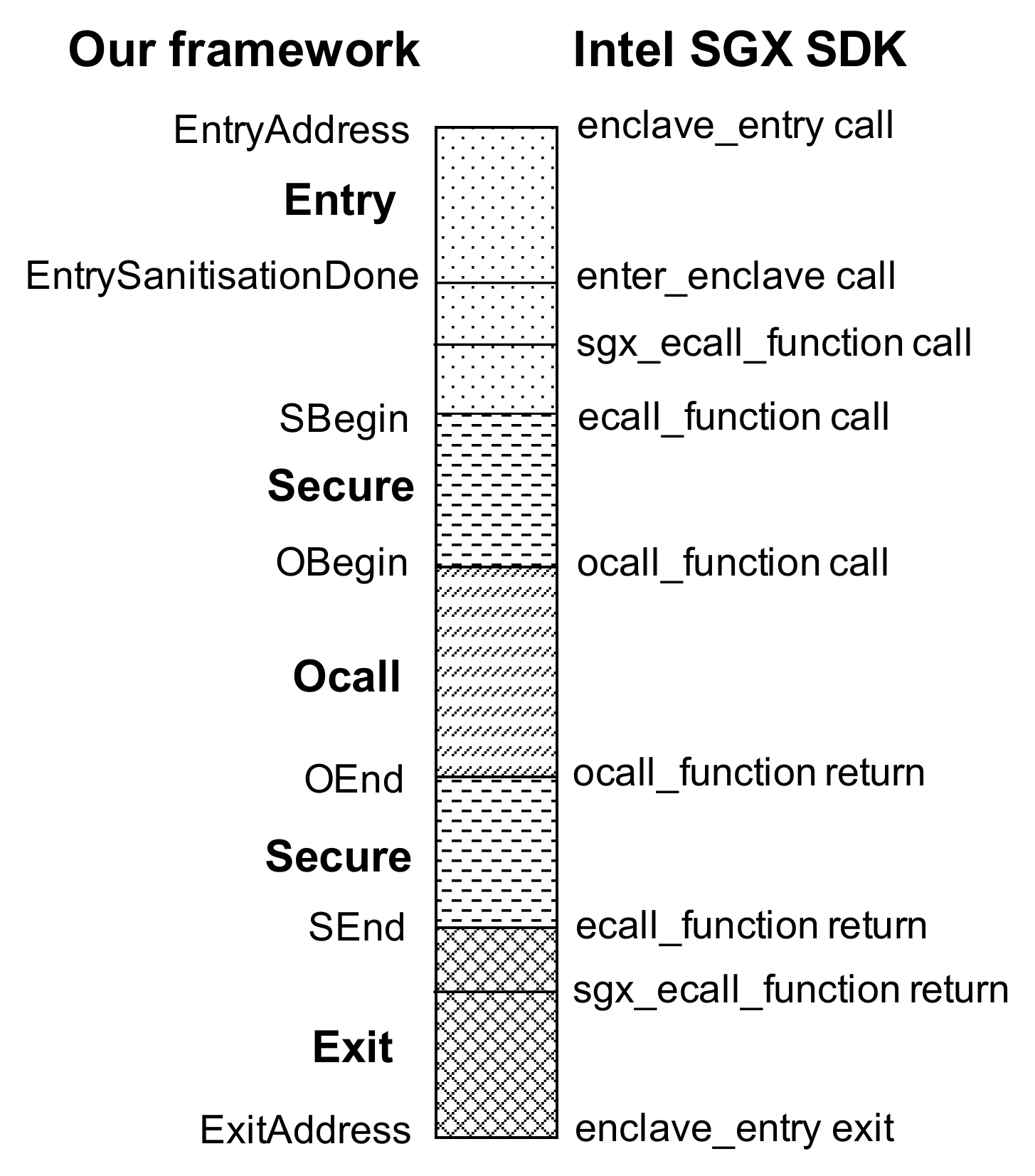}
\end{minipage}
\end{minipage} \hfill
\begin{minipage}[!t]{0.48\textwidth}
\begin{lstlisting}[basicstyle=\scriptsize\sffamily,breaklines=true,caption=Templates for the main functions of an enclave in the Intel SGX SDK programming model.,label=lst:sdk-enclave]
// ecall bridge function edger8r/!*-*!/generated based on our EDL
sgx_status_t sgx_ecall_function(int* i, int* retv)
{
	/!* \lannotation{lst:sgx-ecall-check} *!/ Argument/!*-*!/pointers/!*-*!/outside/!*-*!/enclave check
	/!* \lannotation{lst:sgx-ecall-copy-in} *!/ Copy from untrusted to trusted memory
	ecall_function(t_i, t_retv);
	/!* \lannotation{lst:sgx-ecall-copy-out} *!/ Copy from trusted to untrusted memory
}
// user/!*-*!/created ecall function
void ecall_function(int* i, int* retv) 
{
	/!* \lannotation{lst:ecall-pre-cmds} *!/ secure commands pre-ocall
	ocall_function(v);
	/!* \lannotation{lst:ecall-pos-cmds} *!/ secure command pos-ocall
}
// ocall proxy function edger8r/!*-*!/generated based on our EDL
sgx_status_t ocall_function(int* v)
{
	/!* \lannotation{lst:ocall-check} *!/ Argument/!*-*!/pointers/!*-*!/within/!*-*!/enclave check
	/!* \lannotation{lst:ocall-copy-out} *!/ Copy from trusted memory to untrusted memory
	sgx_ocall(index, u_v)
	/!* \lannotation{lst:ocall-copy-in} *!/ Copy from untrusted memory to trusted memory
}
\end{lstlisting}
\vfill
\end{minipage}
\end{figure}

%Secure
After registers and input memory have been properly handled, the \secure{} phase can start. It captures the proper trusted computation to be carried out. A binary can have many distinct secure functions defined. For instance, in the SDK paradigm, each ecall function typically describes a different secure computation that the enclave implementation offers. In our framework, we use the annotation $\textit{Secure} = (\textit{SBegin}_1, \allowbreak \textit{SEnd}_1),\ldots,\allowbreak (\textit{SBegin}_n,\allowbreak \textit{SEnd}_n)$ to identify these computations where the pair of addresses $(\textit{SBegin}_i, \textit{SEnd}_i)$ identifies the beginning and end of the secure computation $i$. For instance, the address at which \lstinline|ecall_function| is called in \lstinline|sgx_ecall_function| and the address of the instruction following this call (i.e., the return address for this call) could account for such a pair. During the \secure{} phase, the enclave has to follow the \textit{SecurePolicy} which forbids it from reading from and writing to untrusted memory. Performing one of these actions could indicate that the \entry{} phase was not performed correctly and there still is some inputs that need copying into trusted memory, or that some bug exists in the enclave code which is leaking information into untrusted memory.

There are subtleties that should be observed about the EDL-edger8r approach. For instance, the edger8r tool does not deep copy parameters, it only shallow copies pointer elements of a data structure. The developer is in charge of writing specific procedures to carry out deep copies if needed. Moreover, the enclave developer might decide to completely bypass the pointer validation and pointed-to memory copy offered by the EDL-edger8r approach by using the annotation \lstinline|[user_check]|. In this case, the developer should create its own validation and copy code. Note that, for cases where the EDL-edger8r approach is not sufficient or not used, the \secure{} phase will not match exactly the execution of the user-defined ecall function; the code block used for pointer validation and deep copy will be part of this function and the \secure{} phase should start after this block. Including such memory manipulations in this phase would constitute a violation of the \textit{SecurePolicy}.

%Ocalling
During the \secure{} phase, the enclave might need to transition into untrusted code to carry out functionalities that are unavailable within it. Typical examples are OS-assisted functions such as manipulating a file or network socket; system calls are unavailable within the enclave. The SDK's programming model proposes the use of \emph{ocalls} to implement such behaviour. The ocall abstraction is implemented by a \emph{proxy function} in the enclave code that manipulates parameters and call the corresponding ocall function in untrusted code. Since we only consider the enclave's design and code, we are only interested in this proxy function. In Listing~\ref{lst:sdk-enclave}, \lstinline|ocall_function| depicts the template of the proxy function generated by edger8r based on the corresponding EDL definition in Listing~\ref{lst:edl}. Based on the \lstinline|[in,out]| EDL annotation for pointer \lstinline|v|, edger8r generates: a validation that the \lstinline|int| pointed by \lstinline|v| lies within the enclave in~\ref{lst:ocall-check}, the code to copy this \lstinline|int| into untrusted memory in~\ref{lst:ocall-copy-out} (this new copy is pointed to by \lstinline|u_v|), and the code to copy back the \lstinline|int| from \lstinline|u_v| to \lstinline|v| in~\ref{lst:ocall-copy-in}. The call to \lstinline|sgx_ocall(index, l_v)| triggers a series of tRTS functions which are in charge of switching from trusted code into untrusted code and back into enclave code once the untrusted execution is completed, saving and restoring the enclave state in the process. We use the annotation $\textit{OCall} = (\textit{OBegin}_1, \textit{OEnd}_1),\ldots,(\textit{OBegin}_m, \allowbreak \textit{OEnd}_m)$ to identify ocall proxy functions where the pair of addresses $(\textit{OBegin}_i, \allowbreak \textit{OEnd}_i)$ denotes the beginning and end of the \ocall{} phase, respectively. For instance, the address at which \lstinline|ocall_function| is called and the following address in \lstinline|ecall_function| could form such a pair. Multiple ocall functions can be called during the secure execution. For each of them, the enclave transitions from the \secure{} phase to the \ocall{} phase and back. During the \ocall{} phase, the enclave can read and write to untrusted memory; \textit{OcallPolicy} allows these actions. These reads and writes are necessary to pass parameters across the trust boundary.

% Exiting
Finally, after the secure computation has been performed, the enclave transitions into its final phase of execution: \exit{}. At this stage, the enclave has to both sanitise registers to avoid leaking information about the trusted execution, and copy outputs into untrusted memory. Our framework captures this phase with annotation $\textit{Exit} = (ExitAddress)$ where \textit{ExitAddress} marks the end of the enclave execution. This phase starts as soon the \secure{} phase ends and finishes when execution reaches \textit{ExitAddress}. When this address is reached, the predicate \textit{ExitSanitation} must hold, where it expects that registers \emph{rcx}, \emph{rdx}, \emph{r8}-\emph{r15} are cleared, and that stack pointers \emph{rbp} and \emph{rsp} point to untrusted memory. The end of the \lstinline{enclave_entry} assembly routine is in charge of carrying out this sanitation on exit. Moreover, the results of the secure computation, or at least some public part of it, have to be copied out to untrusted memory if they are to be accessed by any entity that is not the enclave itself. The SDK also proposes the EDL-edger8r paradigm to effectively carry out this copy. For the annotation \lstinline|[out]| for \lstinline|ecall_function| parameter \lstinline|retv| in Listing~\ref{lst:edl}, edger8r generates the code to both validate that \lstinline|retv| points to an \lstinline|int| outside the enclave, and to copy the \lstinline|int| pointed by \lstinline|t_retv|, possibly modified by the execution of \lstinline|ecall_function|, into the address pointed to by \lstinline|retv|. This phase enforces \textit{ExitPolicy}: it restricts access to untrusted memory to writes only.

Figure~\ref{fig:execution} presents a typical execution of an orderly enclave with our address annotations and their corresponding SDK commands. Each phase has its own pattern; we label each of them in bold font. Address annotations and SDK commands are in normal font. 

\begin{definition}\label{def:orderly} An enclave is \emph{orderly} with respect to a given \textit{TransitionAnnotations} if and only if the following conditions hold:
\begin{itemize}
	\item \emph{Transition conditions}: the enclave transitions must respect the following rules. It must start at \entry{}. From \entry{}, it can transition to either \secure{} or \exit{}. The \textit{EntrySanitisationDone} address must be reached before the enclave enters the \secure{} phase. From \secure{}, it can transition to either \ocall{} or \exit{}. From \ocall{}, the enclave can only transition back to \secure{}.
	\item \emph{Sanitisation conditions}: at the \textit{EntrySanitisationDone} address, the \textit{EntrySanitisationValidation} predicate must hold, whereas at \textit{ExitAddress}, the \textit{ExitSanitisationValidation} predicate must hold.
	\item \emph{Untrusted memory access policies}: during each of these phases, the enclave must respect the corresponding policy to access untrusted memory.
\end{itemize}
\end{definition}

Satisfying the conditions for being orderly does not guarantee security or correctness, but we assert that following the discipline of orderliness is always a sensible decision when creating enclaves.

\subsection{Enclave analysis with \emph{\Analyser{}}}
\label{sec:analyser}

We introduce \Analyser{}: a tool, built on top of the binary analysis framework \emph{angr}~\cite{Shoshitaishvili16}, that uses symbolic execution to validate an enclave binary against our orderly definition and to look for some typical memory-corruption vulnerabilities.

% How it works: what are the detectors
In \emph{angr}, one can create a Python function to be executed when some specific address is reached in the symbolic exploration of the binary---these functions are called \emph{SimProcedures} and the process of assigning a function to a specific address is called \emph{hooking}. Moreover, one can extend the state of the symbolic execution with some variables described in what is called a \emph{SimState plugin}. Our tool extends the state of the symbolic execution with a variable that keeps track of the different phases and a flag that denotes whether entry sanitisation has occured. We use SimProcedures to check and implement phase transitions, and they are hooked to the corresponding addresses in \emph{TransitionAnnotations}. For instance, SimProcedure \textit{ToSecure} checks whether the enclave is in either \entry{} or \ocall{} phases and whether the entry sanitisation flag is set. If these checks succeed, the enclave execution state is switched to the \secure{} phase. This procedure is hooked at addresses $\textit{OEnd}_i$ and  $\textit{SBegin}_i$. A \emph{transition violation} is reported if a transition is being triggered from the wrong phase, whereas a \emph{sanitisation violation} is reported when some sanitisation condition is not met. We also create SimProcedures to simulate some SGX-specific instructions unrecognised by \emph{angr}.

% Bounds checking and adversarial behaviour
Given the input binary and sizes to the enclave's stack and heap, we determine the enclave's memory layout and instrument data structure \lstinline|global_data_t| with the enclave's size, and its heap's size and start offset with respect to the enclave's base address, and
the \lstinline|thread_data_t| structure with the stack's size and offset with respect to the corresponding Thread Control Structure (TCS); we only consider single-threaded executions. This instrumentation allows our tool to precisely execute memory range checks in the binary. The precise identification of the enclave's memory layout is a key element in our framework. It is used to enforce our policies and stack sanitisation conditions, and to implement some abstractions to account for adversarial behaviour. We use \emph{angr}'s breakpoints to intercept reads and writes to memory and the memory bounds that we calculate to establish whether they satisfy our policies. For instance, a read from an address outside the enclave when in phase \secure{} is reported as a violation. Our tool breaks down such violations into: out-of-enclave read, write, or jump violations. Moreover, our tool ensures that reads from a memory address outside the enclave are symbolic, accounting for adversarial behaviour. Furthermore, we also use breakpoints to report reads from, writes to, and jumps to symbolic addresses within the enclave as they can denote memory-corruption vulnerabilities---we call these \emph{symbolic read, write, and jump violations}, respectively.

% Heuristics & Re-entrant ocalls & Inability to tackle ecalls
We propose a heuristic to automatically generate \textit{TransitionAnnotations} for an input enclave binary to facilitate the use of our tool. It uses debug symbols to create the associations given in Figure~\ref{fig:execution}. Our heuristic should correctly annotate SDK enclaves for which the EDL-edger8r combination works without any adjustment such as tailor-made deep copying. It cannot be used to analyse closed source binaries as their symbols are usually stripped. Moreover, our tool does not account for reentrant ecalls, namely, ecalls triggered within an ocall execution. Hence, violations caused by this sort of behaviour are not reported by our tool. Furthermore, our tool does not account for implicit enclave exit when a hardware exception occurs, even though the corresponding code has been the source of attack vectors~\cite{Biondo18}. Capturing these sorts of behaviour would significantly increase the complexity of our analysis so we left this for future work.

\subsection{Detecting  enclave vulnerabilities}
\label{sec:vulnerabilities}

The main inspiration for our framework is the work in~\cite{Bulck19}. In that work, a number of vulnerabilities are \emph{manually} identified. Our framework's sanitisation conditions and untrusted memory access policies outline tests that can \emph{automatically} identify those vulnerabilities---or their effects. The work in~\cite{Cloosters20} was published as we were creating on our framework and inspired our detection of typical memory-corruption vulnerabilities. In this section, we discuss how our framework can flag behaviours exhibiting vulnerabilities identified in the literature. We split the vulnerabilities into three categories as follows.

\subsubsection{Register sanitisation} As CPU registers are shared between trusted and untrusted execution, they have to be properly sanitised when a transition occurs between these two execution modes. Upon enclave entry, failure to sanitise the the Alignment Check (AC) flag and Direction Flag (DF) in the x86\_64 architecture, for instance, can give an attacker the ability to overly influence and probe the behaviour of an enclave. Moreover, failure to clear registers upon exit can cause information about the trusted execution to leak into the untrusted world. Furthermore, failure to properly switch between trusted and untrusted stack can cause the untrusted environment to both exert undue control over the trusted execution and the leaking of confidential information. Runtime libraries should properly set, restore and clear registers to prevent these vulnerabilities.

Our \textit{EntrySanitisation} and \textit{ExitSanitisation} conditions together with our untrusted access policies can identify such vulnerabilities. For instance, an enclave behaviour that does not clear the AC and DF flags would lead to an execution where \textit{EntrySanitisation} does not hold at address \textit{EntrySanitisationDone}. This execution is a witness for this vulnerability that would be identified by our tool.

\subsubsection{Memory range checking} Failure to properly establish whether some memory range lies within or outside the enclave might lead to vulnerabilities such as writing confidential data into untrusted memory or reading and branching on an attacker-controlled value. The absence of a range check, an overflow on address calculation, or a misuse of SDK functions \lstinline|sgx_is_within_enclave| and \lstinline|sgx_is_outside_enclave| have been identified as causes to such failures. These functions check whether a buffer is completely inside and outside trusted memory, respectively; overlapping trusted-untrusted memory buffers fail both checks. Thus, \lstinline|sgx_is_within_enclave| returning false does not mean that the memory range passed as an argument lies completely outside the enclave---an assumption wrongly made by some developers in designing real enclaves. Another specific instance of lacking or faulty range checks is null-pointer dereferencing. In C/C++ code, the null-pointer value---a pointer to the zero address---is used to convey that a pointer is uninitialised. When a program tries to dereference such a pointer, the process has typically no memory allocated at that address, causing a segmentation-fault crash. Pointer range validation is usually neglected for this address because of that. In the SGX trust model, however, an attacker is able to manipulate the process' virtual memory and allocate some memory at address zero. So, instead of a crash, the attacker can manipulate the value at this address to force an enclave to possibly execute some undesired behaviour.

The untrusted-memory-access policies enforced by our framework should identify effects of these vulnerabilities. For instance, a faulty range validation in the \entry{} phase is likely to lead to some unintended access to untrusted memory in the \secure{} phase. Our framework would capture this behaviour as a violation to the \textit{SecurePolicy}. The enclave developer can, then, analyse the trace leading to this violation and find out whether a range check is missing or faulty.

\subsubsection{Memory copying across trust boundary} Failure to copy data across the trust boundary might cause the enclave to work on untrusted data, being exposed, for instance, to time-of-check time-of-use attacks. As the attacker controls untrusted memory, the contents of a memory buffer may be changed between checking for some condition and using it in some computation. One cause for such failure is wrongly assuming that the EDL-edger8r approach (deep) copies pointer elements of data structures, or simply overlooking the ability of the attacker to manipulate untrusted memory addresses.

Our policies restricting the access to untrusted memory should capture the effects of such a faulty copy. Moreover, we capture the attacker's power to alter the value at untrusted memory addresses at any point by making all values read from untrusted memory symbolic. Our framework tries to strike a compromise by allowing double fetches of untrusted pointers during \entry{} and \ocall{} phases, but no fetch is allowed in \secure{} and \exit{} phases.

\subsection{Evaluation}
\label{sec:evaluation}

\begin{table}[t]
	\caption{Relevant evaluation results.}
	\label{tab:evaluation}
	\centering
	\begin{tabular}{|c|c|c|c|c|c|c|c|c|c|c|c|}
		\multicolumn{1}{c}{Example} & 
		\multicolumn{1}{p{0.3cm}}{\rotatebox{65}{sanitisation}} & 
		\multicolumn{1}{p{0.3cm}}{\rotatebox{65}{out-of-enclave read}} & 
		\multicolumn{1}{p{0.3cm}}{\rotatebox{65}{symbolic read}} & 
		\multicolumn{1}{p{0.3cm}}{\rotatebox{65}{out-of-enclave write}} & 
		\multicolumn{1}{p{0.3cm}}{\rotatebox{65}{symbolic write}} & 
		\multicolumn{1}{p{0.3cm}}{\rotatebox{65}{out-of-enclave jump}} & 
		\multicolumn{1}{p{0.3cm}}{\rotatebox{65}{symbolic jump}} & 
		\multicolumn{1}{p{0.3cm}}{\rotatebox{65}{\#ecalls}} & 
		\multicolumn{1}{p{0.3cm}}{\rotatebox{65}{\#flagged}} & 
		\multicolumn{1}{p{0.3cm}}{\rotatebox{65}{\#timeout}} & 
		\multicolumn{1}{p{0.3cm}}{\rotatebox{65}{\#stopped}} \\
		\hline
		tls-cli & & \checkmark & & & \checkmark & & & 8 & 2 & 5 & 1 \\
		\hline
		seal & & \checkmark &&& \checkmark&& & 6 & 1 & 0 & 1 \\
		\hline
		http & & \checkmark & \checkmark & & \checkmark & & & 3 & 2 & 0 & 1 \\
		\hline
		crypto & & \checkmark &&& \checkmark && & 7 & 1 & 0 & 3 \\
		\hline
		gmp & \checkmark & \checkmark &&&&&& 7 & 3 & 1 & 4 \\
		\hline
		wolfssl & & & \checkmark &&&&& 22 & 13 & 0 & 0 \\
		\hline
		discovery & \checkmark &  & &  & & & & 7 & 0 & 0 & 0\\
		\hline
	\end{tabular}
\end{table}

We have analysed 15 binaries ranging across 12 different enclave samples. We discuss some interesting violations found. The relevant results of our analyses are given in Table~\ref{tab:evaluation}.\footnote{\Analyser{} is open source and available at {\footnotesize\url{https://github.com/blockhousetech/guardian}} together with the enclave binaries we generated.} We report the type of violations that we found for each binary, their number of ecalls, how many exhibited some violation, how many timed out, how many were stopped. We set a 20-minute time budget per ecall analysis. We stop examining an ecall if it reaches a stage where more than 100 branches are being simultaneously explored or 20 violations have been found. The first stop condition is meant to prevent branch explosion whereas the second an excessive number of violations from being reported. We used a machine with Intel i7-9750H CPU, 16GB of RAM, running Ubuntu 18.04.

We analysed four SGX architectural enclaves created by Intel: launch, quoting, provisioning, and provisioning certificate enclaves. We used the version 2.12 of the SDK to build the code at \cite{ae}. Each of them plays a vital role in the SGX trusted ecosystem. These enclaves offer 11 ecalls altogether. Our analysis could not identify any violations on them. As no violations or timeouts were found, we did not add these result to Table~\ref{tab:evaluation}. The absence of violations can be explained by the level of scrutiny these enclaves must have gone through given their importance in Intel's trust ecosystem. Also, it highlights the ability of our tool in checking enclaves without reporting false positives. 

We analysed the enclaves examined by \emph{TeeRex} in~\cite{Cloosters20}; their binaries can be found at~\cite{teerex}. All these enclaves do not properly sanitise AC/DF flags. So, they can be susceptible to the attacks in~\cite{Bulck19}. Not only these enclaves, but any enclave built with SDK versions <2.7.1 will suffer with this vulnerability. For such enclaves, our tool reports an entry sanitisation violation. TeeRex does not check for this sort of sanitisation condition and it does not analyse runtime-library code, and so it is unable to find those. We did not carry out a more thorough analysis of these enclave as they rely on \lstinline|[user_check]|-annotated parameters, and therefore would require manual annotation as opposed to relying on our heuristics. Therefore, these examples are not part of our results table.

Enclaves \textit{tls-cli}, \textit{seal}, \textit{http}, \textit{crypto} use the \textit{Rust SGX SDK} framework~\cite{Wang19}: it ports the Intel SDK to the memory-safe language Rust---we used the version 2.12 of the Intel SDK to build the code at~\cite{teaclave}. \textit{tls-cli} implements a TLS client, \textit{seal} data sealing functions (i.e. data encryption for persistent storage), \textit{http} a http requester, and \textit{crypto} some cryptographic functions. Enclave \textit{gmp} is demo program for the GNU multi precision arithmetic library ported to SGX; we used the version 2.6 of the SDK to build the code at~\cite{gmp}. It offers a few basic mathematical operations and $\pi$ estimation. The \textit{wolfssl} enclave can be used to create a trusted TLS client/server application based on the wolfssl library---we used the version 2.12 of the SDK to build the code at~\cite{wolfssl}. Enclave \textit{discovery} is part of a micro-service implementing private contact discovery; we used the version 2.1.3 of the SDK to build the code at~\cite{signal}. It allows clients to discover which of their contacts are registered for the service without revealing their contacts to the service operator.

For this evaluation, we have added a flag to our tool that indicates whether the enclave under analysis was built using versions 2.1.3 or 2.6, as opposed to version 2.12 of the Intel SDK, which is the version targeted by \Analyser{}. When this flag is set, the sanitisation check is disabled and a slight different instrumentation of the binary is made---these older SDKs use a different data structures to capture the memory layout of an enclave. Moreover, we had to create a new heuristic to identify ocalls in \textit{Rust SGX SDK} enclaves; we added a new flag to enable it for the corresponding examples.

We have noticed a recurrent violation pattern across these enclaves. For an ecall pointer parameter annotated with \lstinline|[in]|, if the corresponding argument for the EDL-edger8r generated bridge ecall function is \lstinline|NULL|, the copy-into-trusted-memory code is not triggered and the null-pointer is passed to the ecall function. Therefore, the ecall function operates on a pointer to typically attacker controlled memory buffer. Note that a dereference to this pointer, in the \secure{} phase, triggers an out-of-enclave-read violation which is detected by our tool. A simple approach to correct such behaviour is to abort the behaviour of an ecall if null pointer is passed to the ecall function.

All \textit{Rust SGX SDK} enclaves have this issue on a common ecall \lstinline|t_global_init_ecall|. This ecall is used to initialise some enclave metadata information: an identifier and a path. Even though the \lstinline|path| parameter is \lstinline|[in]|-annotated, it is afflicted by the violation above and so \lstinline|path| can be made attacker controlled. Parameters \lstinline|cert| (a path to a TLS certificate) and \lstinline|hostname| of \textit{tls-cli}'s ecall \lstinline|tls_client_new|, parameter \lstinline|hostname| of \textit{http}'s ecall \lstinline|send_http_request|, and parameters \lstinline|str_a| and \lstinline|str_a| of \textit{sgx-gmp}'s ecalls \lstinline|e_mpz_add|, \lstinline|e_mpz_mul|, \lstinline|e_mpz_div|, \lstinline|e_mpf_div| can all be similarly subverted by passing a null pointer. This violation gives the attacker the power to change the value of these arguments even after the bridge function has finished processing inputs. We have contacted maintainers of these projects who have agreed these were issues and have already fixed them.\footnote{\scriptsize\url{https://github.com/apache/incubator-teaclave-sgx-sdk/pull/322},\linebreak\url{https://github.com/intel/sgx-gmp-demo/commit/a05da606b2cfd4710c80e3c99068a6b97dc31888}} We have tested these enclaves after they have been fixed, and we can confirm that \Analyser{} no longer finds these violations. As some of these enclaves had already been analysed by TeeRex and these vulnerabilities had not been reported, we have evidence that TeeRex was unable (or did not consider) these behaviours vulnerabilities---TeeRex is not publicly available for use.

Based on a fundamentally different philosophy than similar enclave analysis techniques, our tool is capable of identifying subtle violations in a reasonable time frame. The most relevant violations that we identified were not detected via the typical memory-corruption vulnerabilities proposed by TeeRex but via our policies on untrusted memory accesses and sanitisation conditions.

\section{Conclusions}
\label{sec:conclusion}

% Highlight contribution
We propose a definition of orderly enclave as a means to capture the intent of enclave developers; we define where phase transitions occur and, consequently, how enclaves should behave at different stages of their execution. Based upon this notion, we created \Analyser{}: a tool that uses symbolic execution to validate enclaves' binaries against this definition---and also to find some typical memory-corruption vulnerabilities. Violations to this definition uncovered by \Analyser{} are likely to expose vulnerabilities arising from improper interactions between the trusted and untrusted resources. We also demonstrate that our framework is capable of efficiently identifying violations in practical enclaves, uncovering in some case subtle behaviours. We identify violations in practical enclaves which went undetected when analysed by similar state-of-the-art tools. These results show that \Analyser{} can be a valuable tool in improving the security of enclaves, helping developers to be more confident about their design and implementation.

We plan to add some coverage metrics and to experiment with fine-grained notions of confidentiality. Another topic for future work is the feasibility of automatic binary hardening for the violations that we find. 

\bibliographystyle{splncs04}
\bibliography{reference}

\end{document}